\documentclass[aps,prl,twocolumn,groupedaddress,amsmath,amssymb,lengthcheck]{revtex4}
\usepackage{graphicx}% Include figure files
\usepackage{dcolumn}% Align table columns on decimal point
\usepackage{bm}% bold math
\usepackage{hyperref}% add hypertext capabilities
\usepackage[caption=false]{subfig}

\begin{document}

%\preprint{}

\title{Thermocapillary Flow on Superhydrophobic Surfaces}


\author{Tobias Baier}
\email[]{baier@csi.tu-darmstadt.de}
%\homepage[]{www.csi.tu-darmstadt.de}
%\thanks{}
\affiliation{Technische Universit\"at Darmstadt, Center of Smart Interfaces, Petersenstra\ss e 32, 64287 Darmstadt, Germany}

\author{Clarissa Steffes}
\affiliation{Technische Universit\"at Darmstadt, Center of Smart Interfaces, Petersenstra\ss e 32, 64287 Darmstadt, Germany}

\author{Steffen Hardt}
\affiliation{Technische Universit\"at Darmstadt, Center of Smart Interfaces, Petersenstra\ss e 32, 64287 Darmstadt, Germany}

\date{\today}

\begin{abstract}
A liquid in Cassie-Baxter state above a structured superhydrophobic surface is ideally suited for surface driven transport due to its large free surface fraction in close contact to a solid. We investigate thermal Marangoni flow over a superhydrophobic array of fins oriented parallel or perpendicular to an applied temperature gradient. In the Stokes limit we derive an analytical expression for the bulk flow velocity above the surface and compare it with numerical solutions of the Navier-Stokes equation. Even for moderate temperature gradients comparatively large flow velocities are induced, suggesting to utilize this principle for microfluidic pumping.
%The technique constitutes a new method for pumping in microfluidic settings.
\end{abstract}

%\keywords{}

\maketitle


%\section{Introduction}
{\it Introduction --} Microtextured surfaces have mainly received attention due to their wetting properties \cite{quere2008}. A liquid drop placed on a suitably structured hydrophobic surface will only be in contact with the material on protruding tips, while gas is trapped in the valleys in between. In this so called Cassie-Baxter state nearly perfect hydrophobicity can be obtained, reflected in contact angles close to $180^\circ$. Recently, such surfaces have gained interest with respect to their ability for drag reduction \cite{ybert2007,rothstein2010} and surface induced transport \cite{huang2008,ajdari2006}, in particular electroosmotic and diffusioosmotic flow.

\begin{figure}[b]
\includegraphics[width=0.7\columnwidth]{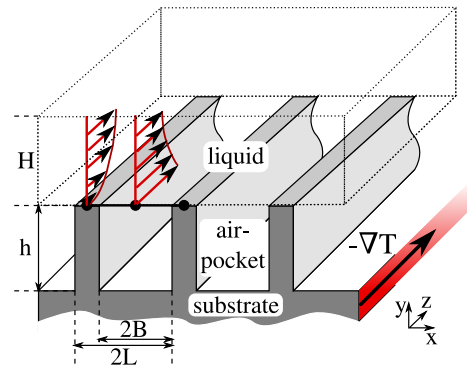}
\caption{\label{fig:sketch} (color online) Sketch of the geometry. The $z$-axis is always chosen along the main temperature gradient, the $y$-axis always normal to the structured surface. In this example we assume a temperature gradient along the fins (longitudinal).}
\end{figure}

In this letter we analyze temperature induced Marangoni convection as a driving force for fluid transport along microtextured surfaces. In particular, we focus on finned surfaces as sketched in figure \ref{fig:sketch}, with a temperature gradient along or perpendicular to the fins, and the liquid being in the Cassie-Baxter state. We use an integral relation for the Stokes equation to derive an analytical formula for the macroscopic flow velocity observed at some distance above the surface. Both situations are further investigated by numerical solutions of the Navier-Stokes equation and compared to the analytical formula. Our analysis focuses on substrates of high thermal conductivity, such as silicon.

Fluid actuation and transport are core functionalities in many microfluidic systems. The most prominent examples for the corresponding driving mechanisms are pressure-driven and electroosmotic flow. Our analysis shows that moderate temperature gradients of the order of $10\; \text{K/cm}$ can lead to fluid velocities of several $\text{mm/s}$ for water based systems on superhydrophobic surfaces. Thermocapillary convection may thus add to the portfolio of actuation principles in microfluidic settings and may even enable larger flow velocities than typically achieved with electroosmosis.

{\it Marangoni flows --} The stress on a liquid-gas interface due to a gradient in surface tension is \cite{landau1987}
\begin{equation}
  \label{eq:marangoni}
  n_i \tau_{ij} t_j= - t_i \partial_i \sigma,
\end{equation}
where $n_i$ and $t_i$ are components of the interface normal and tangential vectors in $i$ direction, and we use the convention to sum over repeated indices. $\tau_{ij}$ are the components of the stress tensor and $\sigma$ is the surface tension. For an incompressible Newtonian fluid characterized by the viscosity $\eta$ the total stress tensor is
\begin{equation}
  \label{eq:stresstensor}
  \tau_{ij}=\eta \left(\partial_i u_j+\partial_j u_i \right) - p\delta_{ij},
\end{equation}
consisting of a viscous part proportional to the shear rate tensor and a part originating from the pressure field $p$. The equation of motion for the fluid is the stationary Navier-Stokes equation
\begin{equation}
\label{eq:stokes}
  \partial_j (\rho u_j u_i )=\partial_j \tau_{ji}, \hspace{0.5cm}
  \partial_i u_i = 0,
\end{equation}
where the fluid is assumed to be incompressible. The left hand side of the momentum equation becomes negligible at small velocities (Reynolds numbers), and we will call this limit the Stokes limit, with the corresponding equation of motion being the Stokes equation. On the fin surface we assume a no-slip boundary condition, while eq. (\ref{eq:marangoni}) constitutes the corresponding boundary condition on the liquid-gas interface.

The temperature in the fluid is governed by the stationary limit of the energy equation
\begin{equation}
\label{eq:T2D}
\rho c_p u_i \partial_i T = k\Delta T,
\end{equation}
where $\rho$, $c_p$ and $k$ denote the density, specific heat capacity and thermal conductivity of the liquid, respectively. In order to tackle the problem analytically, we assume the density, viscosity, heat capacity and thermal conductivity to be independent of temperature while linearizing the temperature dependence of the surface tension \cite{vargaftik1983}. As long as the temperature differences do not become too large this is a suitable approximation. Moreover, we will restrict our analysis to a flat liquid-gas interface.

Equating the viscous part of the stress tensor, $\sim\eta U/L$, with the Marangoni stresses, $\sim\frac{\partial\sigma}{\partial T}  \partial_z T$, yields a dimensionless number, $\beta=\frac{\eta U/L}{\frac{\partial\sigma}{\partial T}  \partial_z T}$, which for a given geometry characterizes the flow problem together with the Reynolds number, $Re=\rho UL/\eta$, and Prandtl number, $Pr=c_p \eta/k$. Below, we will demonstrate that $\beta$ is closely related to the macroscopic slip length in a simple shear flow over textured geometries.

{\it Longitudinal fins --} Consider a geometry where the temperature gradient in the bulk of the substrate has the value $\langle\partial_z T\rangle$ and is parallel to the fins, which defines the $z$-direction (figure \ref{fig:sketch}). The geometry is implied to be of infinite extent in the plane of the substrate and at a height $H$ above the substrate a symmetry plane is assumed. In that case there is translational symmetry in $z$-direction and hence the $z$-component of the temperature gradient will have the same value everywhere in the geometry. Moreover, the velocity and pressure in the fluid will not depend on this coordinate. This simplifies the energy equation (\ref{eq:T2D}) to a 2D equation
\begin{equation}
  \label{eq:T2D_long}
  \rho c_p (\bm{u}_\bot \bm{\nabla}_\bot T + u_z \langle\partial_z T\rangle)=k\Delta_\bot T,
\end{equation}
where we have used the shorthand notation $\bm{u}_\bot \bm{\nabla}_\bot=u_x \partial_x + u_y \partial_y$ and $\Delta_\bot=\partial_x^2+\partial_y^2$. The stationary Navier-Stokes equations partially decouple into a 2D perpendicular equation for $\bm{u}_\bot=(u_x,u_y)$ and a convection diffusion equation for the longitudinal velocity $u_z$:
\begin{align}
  \rho(\bm{u}_\bot \bm{\nabla}_\bot ) \bm{u}_\bot&=\eta \Delta_\bot \bm{u}_\bot - \nabla_\bot p, \nonumber\\
\label{eq:u2D_long}
  \rho(\bm{u}_\bot \bm{\nabla}_\bot ) u_z &= \eta\Delta_\bot u_z,\\
  \partial_x u_x+&\partial_y u_y=0. \nonumber
\end{align}
In the Stokes limit the left hand sides can be set equal to zero and the equation for $u_z$ decouples from the perpendicular parts.

As noted above, the $z$-component of the temperature gradient along the fins is constant everywhere due to the translational symmetry and this leads to a main flow along the fins towards the colder regions. Additionally, temperature variations perpendicular to the main flow occur, the temperature being lower close to a fin than in the middle between two fins. Due to the symmetry of the problem, these perpendicular temperature gradients will only lead to perpendicular vortices and not produce any net flow. %This is in close analogy to electroosmotic flow in geometries with varying surface charge investigated in the stokes limit by Ajdari \cite{ajdari1996}, who finds that no net fluid motion occurs over surfaces having no total charge (corresponding to no net shear stress).

{\it Lorentz reciprocal theorem: --} This integral relation for the Stokes equation \cite{lorentz1896} has in microfluidic applications already proven useful in the case of electroosmotic flow \cite{squires2008}. It is obtained by assuming two solutions for the Stokes equation (\ref{eq:stokes}), one characterized by "unhatted" fields $(\bm{u},p)$ and the other denoted by the respective "hatted" values $(\hat{\bm{u}},\hat{p})$ corresponding to some reference flow field, possibly subject to different boundary conditions. Multiplying equation (\ref{eq:stokes}) in the Stokes limit by $\hat{\bm{u}}$, integrating over a control volume $\Omega$, integrating by parts and subtracting the corresponding equation where "hatted" and "unhatted" fields are interchanged, leaves only the boundary integrals
\begin{equation}
\label{eq:lorentzU}
  \eta \int_{\partial\Omega} dA_i (\hat{u}_j \partial_i u_j - u_j \partial_i \hat{u}_j) = \int_{\partial\Omega} dA_i (\hat{u}_i p - u_i \hat{p}),
\end{equation}
where the continuity equation was used to simplify the pressure term.

Let us now take as the reference flow a Couette flow over the fin geometry, fulfilling the no-slip boundary condition at the solid-liquid and a zero shear stress boundary condition at the liquid-gas parts of the interface. Far away from the substrate a constant shear rate $\hat\gamma$ is assumed. This flow has been studied by Philip \cite{philip1972}, who gives an analytical formula for the flow field and the corresponding macroscopic slip velocity, expressing that far away from the substrate it appears as if the liquid glides over the microstructured surface. However, we shall not need the details of this flow field but only the "global" parameter, the slip length $\hat\beta_l$ over the surface. In particular, the flow field far away from the surface has the form
\begin{equation}
\label{eq:couette}
  \hat{\bm{u}}\sim \hat{\gamma}(y+\hat\beta_l) \bm{e}_z.
\end{equation}

As our control volume $\Omega$ we take a rectangular box spanning the distance from the center of some fin to the center of its neighbor, stretching an arbitrary distance along the fins and being sufficiently high, such that the flow field on the top surface is unaffected by the details induced by the bottom surface, the plane containing the solid-liquid and liquid-gas interfaces. Assuming the top surface to be located at a distance $H$ away from the bottom surface we find that at this position the Couette flow is approximately given by equation (\ref{eq:couette}), while for the ("unhatted") thermocapillary flow a vanishing shear stress can be assumed, since the driving force is surface tension. This leads to a constant flow velocity at at the top surface, i.e.
\begin{equation}
 \label{eq:plug}
  {\bm u} \sim u_{th,l} \bm{e}_z.
\end{equation}

In the surface integrals of eq. (\ref{eq:lorentzU}) we start by considering the "front and back" surfaces perpendicular to the fins. Due to the translational symmetry along the fins and the vanishing of all gradients in this direction the integrals originating from the shear-stress contribution vanish and those originating from the pressure contribution cancel when summing over the front and back surfaces. The same applies to the surfaces parallel to the fins, since these faces constitute symmetry planes. Furthermore, since the flow is parallel to the top and bottom faces the integrals containing the pressure vanish, and we are left with the shear-integrals over these surfaces. From eqns. (\ref{eq:couette}) and (\ref{eq:plug}) we find for the top surface
\begin{equation}
\eta \int_{top} dA_i (\hat{u}_j \partial_i u_j - u_j \partial_i \hat u_j) = -\eta u_{th,l} \hat\gamma A,
\end{equation}
where $A$ is the area of the surface. The integral over the bottom surface only has contributions from the liquid-gas interface since the velocity vanishes on the fins
\begin{equation}
\eta \int_{bottom} dA_i (\hat u_j \partial_i u_j-u_j \partial_i \hat u_j)
  =\partial_z \sigma \int_{liquid-gas} dA_y \hat u_z,
\label{eq:evalLorentz}
\end{equation}
where we have used the fact that the reference flow only has components in $z$-direction and a vanishing shear rate at the liquid-gas interface. We thus arrive at
\begin{equation}
\label{eq:MarangoniVelocity}
  u_{th,l}=\frac{\partial_z \sigma}{\eta\hat\gamma}\frac{1}{A} \int_{liquid-gas} dA_y \hat u_z
  = \frac{\partial_z \sigma}{\eta}\hat\beta_l,
\end{equation}
where the last equality comes from another application of the Lorentz reciprocal theorem taking a Couette flow with no-slip boundary condition everywhere on the bottom surface and shear rate $\hat\gamma$ as "unhatted" flow, i.e. $\bm{u}=\hat\gamma y \bm{e}_z$, while retaining the Couette flow over longitudinal fins as reference flow. Also here only the integrals on the top and bottom surface containing the shear rate contribute, giving $\eta\hat\gamma^2\hat\beta_l A$ on the top surface and $\eta\hat\gamma\int_{liquid-gas} dA_y \hat u_z$ at the bottom.

Equation (\ref{eq:MarangoniVelocity}) is our result for the longitudinal thermocapillary velocity in the Stokes limit. The corresponding longitudinal slip length is \cite{philip1972}
\begin{equation}
\label{eq:beta_l}
  \beta_l = -\frac{2}{\pi}\ln\cos\frac{\pi a}{2},
\end{equation}
where $a=B/L$ is the liquid-gas fraction of the surface and $\beta_l=\hat\beta_l/L$ the dimensionless slip length in terms of half the fin spacing (c.f. figure \ref{fig:sketch}). For later comparison, we invert relation (\ref{eq:MarangoniVelocity}) and define the dimensionless "longitudinal thermocapillary slip coefficient" as
\begin{equation}
\label{eq:beta_Ml}
  \beta_{th,l}=u_{th,l} \frac{\eta/L}{\frac{\partial \sigma}{\partial T} \langle\partial_z T\rangle}.
\end{equation}

\begin{table*}[!htb]%The best place to locate the table environment is directly after its first reference in text
\caption{
\label{tab:BC}
Boundary conditions and model parameters. The corresponding geometry is displayed in figure \ref{fig:beta_M}.
}
\begin{ruledtabular}
\begin{tabular}{cll}
\textrm{boundary}&
\textrm{longitudinal}&
\textrm{transverse}\\
\colrule
1, $1A$, $1B$ & $u_x=u_y=u_z=0,\;T=T_0$ & $u_y=u_z=0$,\; $T|_{1A}=T_0,\; T|_{1B}=T_0 + 2L\langle\partial_z T\rangle$\\
2 & $\tau_{xy}=-\frac{\partial \sigma}{\partial T} \partial_x T,\; \tau_{yz}=-\frac{\partial \sigma}{\partial T} 		  \langle\partial_z T\rangle,\; u_y=0,\; \partial_y T=0$
  & $\tau_{yz}=-\frac{\partial \sigma}{\partial T} \langle\partial_z T\rangle,\; u_y=0,\;\partial_y T=0$\\
3, $3A$, $3B$
  & $u_x=0,\; \tau_{xy}=0,\; \partial_x u_z=0,\; \partial_x T=0$
  & $u_i |_{3A}=u_i |_{3B},\; p|_{3A}=p|_{3B},\; \tau_{yz}=0,$\\
&& $T|_{3B}=T|_{3A}+2L\langle\partial_z T\rangle,\; \partial_z T|_{3B}=\partial_z T|_{3B}$\\
4 & $u_y=0,\; \tau_{xy}=0,\; \partial_y u_z=0,\; \partial_y T=0$ & $u_y=0,\; \tau_{yz}=0,\; \partial_y T=0$\\
\hline\hline
\multicolumn{3}{c}{$\eta=1\,\text{mPa\,s}$, $\rho=1000\,\mathrm{kg/m^3}$, $k=0.6\,\text{W/(m\,K)}$,  $c_p=4200\,\text{J/(kg\,K)}$, $\frac{\partial \sigma}{\partial T}=-0.155\,\text{mN/(m\,K)}$, $T_0=300\,\text{K}$}
\end{tabular}
\end{ruledtabular}
\end{table*}

For realistic values of $10\,\mathrm{\mu m}$ wide fins with $40\,\mathrm{\mu m}$ spacing in between and an applied temperature gradient of $\langle\partial_z T\rangle = -10\, \text{K/cm}$ we obtain from equations (\ref{eq:beta_l}) and (\ref{eq:MarangoniVelocity}), using the values of table \ref{tab:BC}
\begin{equation}
\label{eq:parameters}
  \beta_l=0.7476, \, \hat\beta_l = 18.7\,\mathrm{\mu m}, u_{th,l} = 2.9\,\text{mm/s}.
\end{equation}
Note that this corresponds to a Reynolds number of approximately $Re\sim 0.07$, so the Stokes equation is still expected to be a good approximation. %This velocity is orders of magnitude above values achievable with electroosmotic flow.

In the derivation above we assumed that the top surface of our control volume is located sufficiently high above the structured surface such that the thermocapillary flow field corresponds to a core plug flow. In fact, our numerical calculations confirm that this assumption is fulfilled reasonably well already at a distance of $2L$.

{\it Transverse fins --} In the case of thermocapillary convection perpendicular to the fins, the temperature gradient on the liquid-gas interface is not constant in the main flow direction and it is not possible to pull $\partial_z \sigma$ out of the integral, as done in equation (\ref{eq:evalLorentz}). In fact, only in the case where the Marangoni stresses on the liquid-gas interface can be decomposed into a constant (leading to a main flow) and a part antisymmetric with respect to reflection at a fin center (contributing only to convective rolls but no main flow) can the Lorentz reciprocal theorem be applied as before. Nevertheless, we expect the transverse slip coefficient $\beta_t=\beta_l/2$ obtained for Couette flow over a finned geometry \cite{philip1972} to define an upper bound for the thermocapillary slip velocity that can be obtained. In analogy to equation (\ref{eq:beta_Ml}) we therefore define the "transverse thermocapillary slip coefficient"
\begin{equation}
\label{eq:beta_Mt}
  \beta_{th,t}=u_{th,t}  \frac{\eta a/L}{\frac{\partial \sigma}{\partial T} \langle\partial_z T\rangle}.
\end{equation}
In this definition we have incorporated the fact that the gradient driving the thermocapillary convection is $\langle\partial_z T\rangle_{free}=a^{-1} \langle\partial_z T\rangle$ since the temperature on the solid-liquid interfaces is approximately constant, owing to the assumed high thermal conductivity of the substrate. This seems to contradict the constant temperature gradient assumed in the bulk of the substrate. However, for fins much taller than wide, as usually applied for superhydrophobic surfaces, the diffusional character of the heat conduction equation assures that only the average temperature from the bottom prevails at the top of the fin.

At this point a remark on the constant-temperature boundary condition on the solid-liquid interfaces is in order. This approximation will only be valid if the heat flux from the liquid is small enough. From dimensional considerations the wall heat flux in a fin scales as $J_{fin} \sim \rho c_p u_{th,l} \langle\partial_z T\rangle H/(1-a)$. For an estimate we insert the values used to arrive at equation (\ref{eq:parameters}) together with a geometry height $H=200\,\mathrm{\mu m}$ and obtain as a measure for the temperature gradient in the fin $J_{fin}/k_{Si} \sim 0.8\,\text{K/cm} \ll \langle\partial_z T\rangle$, where the thermal conductivity of silicon $k_{Si}=148\,\text{W/(m\,K)}$ was used. Thus the temperature gradient in the fin is expected to be much smaller than the applied temperature gradient. The largest temperature gradient we will consider is $\langle\partial_z T\rangle=-60\,\text{K/cm}$, which from the above estimate is seen to push our model to its limits. Nevertheless, this only gives a typical size of the temperature gradient in the fin, which will be mostly directed along its height and not along the solid-liquid interface.

\begin{figure*}[!htb]
  (a) \includegraphics[width=0.6\columnwidth]{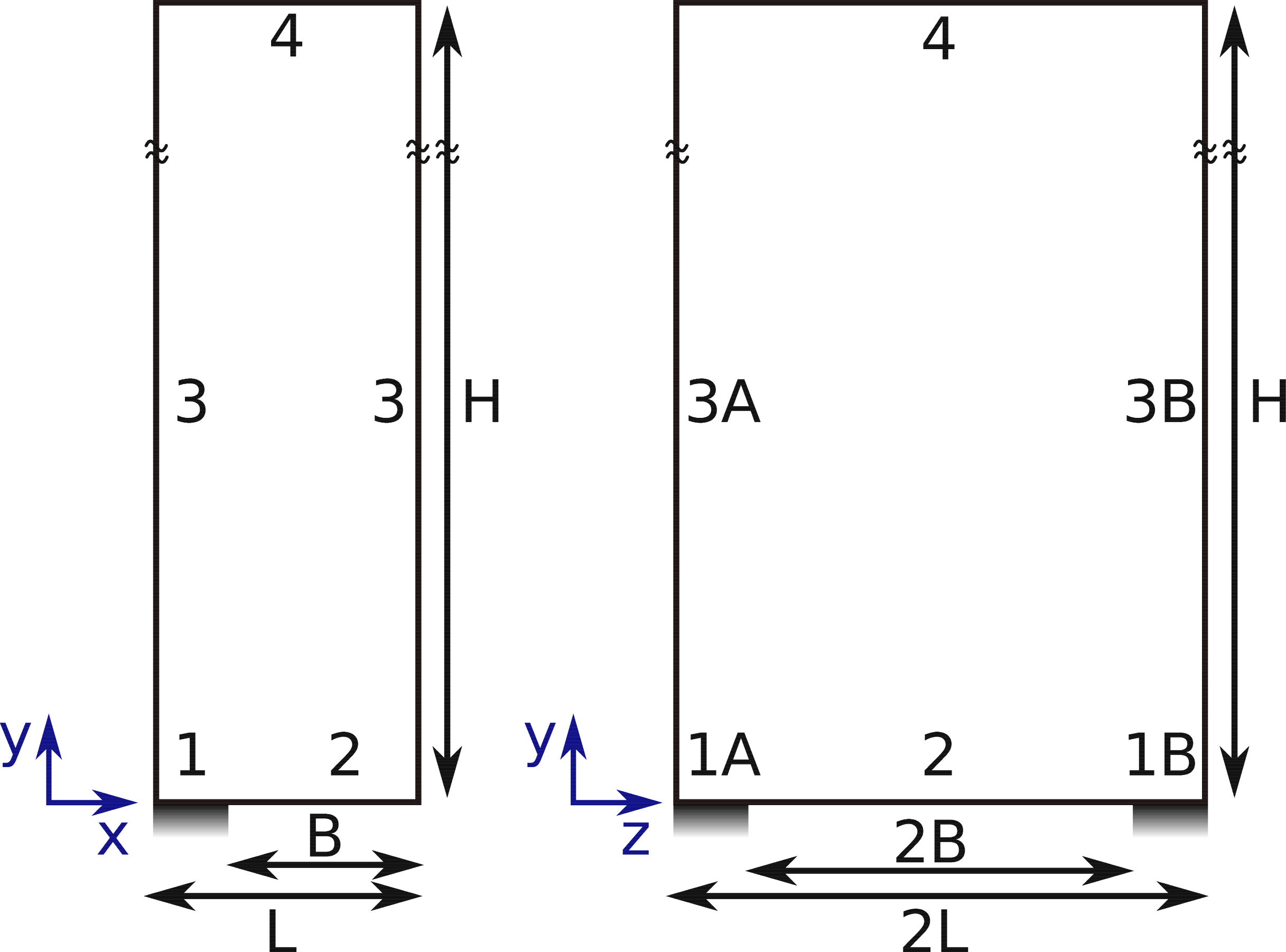}
  \hspace{0.25cm} (b) \hspace{-0.25cm}
  \includegraphics[width=0.65\columnwidth]{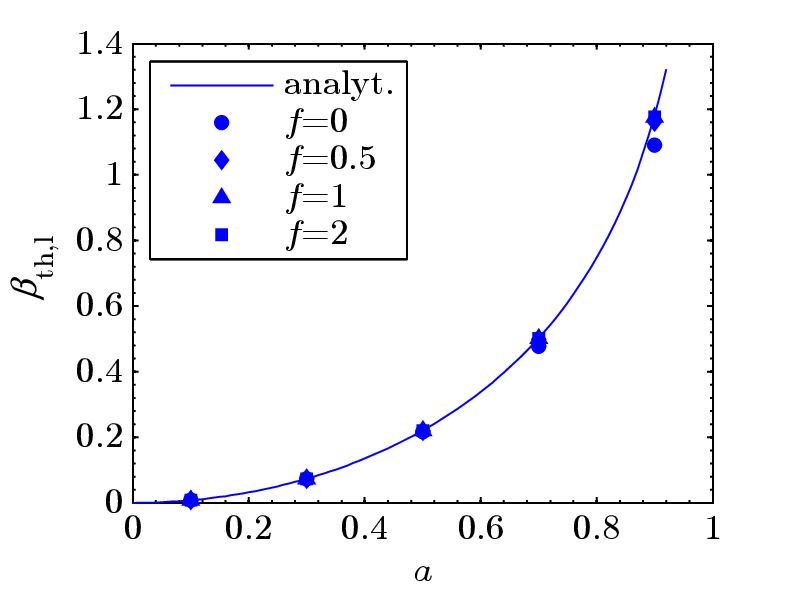}
  \hspace{-0.25cm} (c) \hspace{-0.25cm}
  \includegraphics[width=0.65\columnwidth]{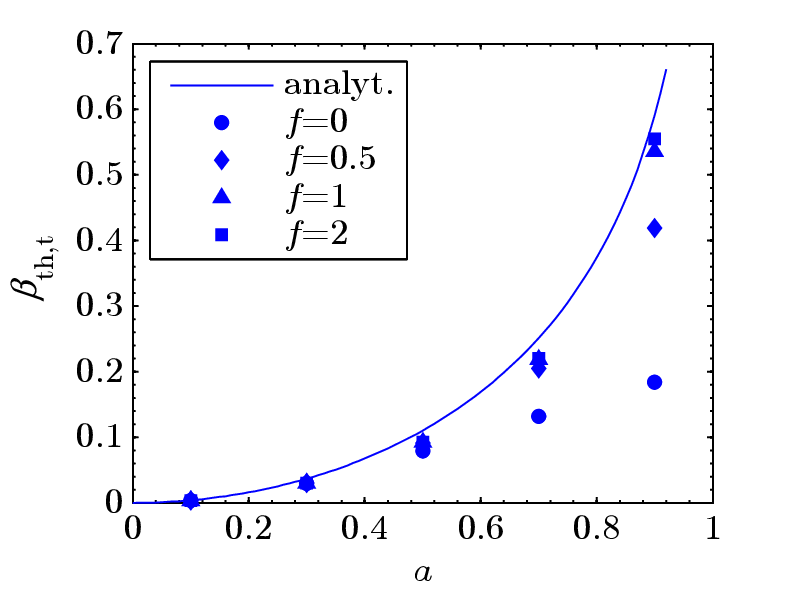}
\caption{(color online) (a) Simulation domains for calculating longitudinal (left) and transverse flow (right). The $z$-axis is always chosen along the main temperature gradient, the $y$-axis always normal to the structured surface. Boundary conditions and model parameters are specified in table \ref{tab:BC}. (b) Longitudinal and (c) transverse thermocapillary slip coefficients as defined in eqs. (\ref{eq:beta_Ml}, \ref{eq:beta_Mt}) for different free surface fractions $a=B/L$ and temperature gradients parametrized by $\langle\partial_z T\rangle=-10^{-f}\;60\,\text{K/cm}$. The solid lines show the respective slip coefficients as calculated by Philip \cite{philip1972}, i.e. eq. (\ref{eq:beta_l}) for the longitudinal case and half of this value for the transverse case.}
\label{fig:beta_M}
\end{figure*}

{\it Numerical results --} To test the analytical result (\ref{eq:MarangoniVelocity}) and to investigate the influence of inertia and non-uniform temperature gradients at the liquid-gas interface we have solved equations (\ref{eq:T2D_long}, \ref{eq:u2D_long}) in the longitudinal case and equations (\ref{eq:stokes}, \ref{eq:T2D}) in the transverse case using a finite element discretization (Comsol mutiphysics). Due to symmetry a 2D calculation suffices in both cases and we restrict our attention to the liquid while assuming the temperature of the substrate as given. The corresponding geometries are shown in figure \ref{fig:beta_M}. For definiteness we consider a fin spacing of $2L=50\,\mathrm{\mu m}$ and a geometry height of $H=200\,\mathrm{\mu m}$ while varying the liquid-gas interface area fraction $a=B/L$. The boundary conditions as well as model parameters are given in table \ref{tab:BC}. In particular, the wall temperature is prescribed, while the liquid-gas interfaces are assumed to be adiabatic.

Figure \ref{fig:beta_M} shows longitudinal and transverse thermocapillary slip coefficients for different free surface fractions and temperature gradients. The solid lines show the respective slip coefficients as calculated by Philip \cite{philip1972}. In the longitudinal case slight deviations from the value obtained in the Stokes limit are observed at temperature gradients above $\langle\partial_z T\rangle=-6 \,\text{K/cm}$. This is expected, since for the largest temperature gradient of $-60\,\text{K/cm}$ the velocities are $\sim25\,\text{mm/s}$ for $a=0.9$, corresponding to $Re \sim 1$. In the transverse case, some larger deviations appear. However, the data points stay below the limiting value obtained from the results of Philip even at low $Re$. Owing to non-constant temperature gradients along the liquid-gas interfaces, the flow velocity is reduced compared to the limiting value, an effect that shows up especially at larger Reynolds numbers. Apparently, with the longitudinal arrangement significantly larger flow velocities can be reached than with the transverse one.

%We remark at this point that a grid independence study shows classical "Richardson convergence" of the results for the thermocapillary slip length towards the theoretical value (\ref{eq:beta_l}) in the Stokes limit, with the difference $|\beta_{th,l}-\beta_l|$ decreasing almost inversely proportional to the grid spacing. From this we estimate the accuracy of the calculated values for $\beta_{th}$ to be of the order of $~3\cdot 10^{-3}$, i.e. far below the marker size in figure \ref{fig:beta_M}.

{\it Conclusion --} We have analytically derived a relation for the thermocapillary flow velocity along a finned superhydrophobic surface. Even at moderate temperature gradients the resulting velocity values are large enough to suggest using this principle for microfluidic pumping. The presented relationships between the slip length and the thermocapillary flow velocity give a very good approximation for the bulk fluid transport in the case of longitudinal fins and represent an upper limit to the flow velocity achievable with transverse fins. We expect that in the same spirit simple relationships for the thermocapillary flow velocity along alternative superhydrophobic surface structures can be found.

\begin{acknowledgments}
We kindly acknowledge support by the German Research Foundation (DFG) through the Cluster of Excellence 259 and grant number HA 2696/12-1.
\end{acknowledgments}

%\nocite{*} %solange noch nix explizit zitiert ist
%\bibliography{literaturprl}
%\input{marangoni.bbl}

%%%%%%%%%%%%%%%%%%%%%%%%%%%%%%%%%%%%%%%%% marangoni.bbl %%%%%%%%%%%%%%%%%%%%%%%%%%%%%%%%%%%%%%%%%%%%%

\end{document}